\documentclass[8pt]{article}  \usepackage{times}
\usepackage{graphicx}

\usepackage{color}

\usepackage[round,numbers,sort&compress]{natbib} 
\begin{document}

\twocolumn[{\LARGE \textbf{Comparing ion conductance recordings of synthetic lipid bilayers with cell membranes containing TRP channels\\*[0.0cm]}}

{\large Katrine R. Laub$^1$, Katja Witschas$^2$, Andreas Blicher$^{1}$,
S\o ren B. Madsen$^1$, Andreas L\"uckhoff$^2$ and \\Thomas Heimburg$^{1, \ast}$\\*[-0.1cm]
{\small $^1$Niels Bohr Institute, University of Copenhagen, Blegdamsvej 17, 2100 Copenhagen \O, Denmark\\*[-0.1cm]}
{\small $^2$Institute of Physiology, Medical Faculty, RWTH Aachen University, D-52057 Aachen, Germany\\*[0.0cm]}

{\normalsize ABSTRACT\hspace{0.5cm} In this article we compare electrical conductance events from single channel recordings of three TRP channel proteins (TRPA1, TRPM2 and TRPM8) expressed in human embryonic kidney cells with channel events recorded on synthetic lipid membranes close to melting transitions.  Ion channels from the TRP family are involved in a variety of sensory processes including thermo- and mechano-reception. Synthetic lipid membranes close to phase transitions display channel-like events that respond to stimuli related to changes in intensive thermodynamic variables such as pressure and temperature.  TRP channel activity is characterized by typical patterns of current events dependent on the type of protein expressed. Synthetic lipid bilayers show a wide spectrum of electrical phenomena that are considered typical for the activity of protein ion channels. We find unitary currents, burst behavior,  flickering, multistep-conductances, and spikes behavior in both preparations.  Moreover, we report conductances and lifetimes for lipid channels as described for protein channels. Non-linear and asymmetric current-voltage relationships are seen in both systems. Without further knowledge of the recording conditions, no easy decision can be made whether short current traces originate from a channel protein or from a pure lipid membrane.\\*[0.0cm] }}
]

\noindent\footnotesize {$^{\ast}$corresponding author, theimbu@nbi.dk --- K. R. Laub and K. Witschas contributed equally to this work.}\\

\footnotesize{\textbf{Abbreviations:} DMPC - 1,2 dimyristoyl-3-sn-phosphatidylcholine; DMPG - 1,2 dimyristoyl-3-sn-phosphatidylglycerol; AITC - allylisothiocyanate ; } 
\footnotesize{pdf - probability distribution function; HEK cell - human embryonic kidney cell; TRP channel - transient receptor potential channel; ADPR - adenosine diphosphate ribose}

\normalsize
\section*{Introduction}
The cell membrane forms a diffusion barrier for hydrophilic substances, and its integrity is crucial for the viability of living organisms. 

The established view is that ion channels selectively regulate the flow of ions across the membrane, permitting the passage of certain ions while excluding others. In addition to their selectivity, protein ion channels have been reported to be functionally characterized by many properties such as conductance, voltage dependence, ligand activation, temperature- and mecha\-nosensitivity.
The synthetic lipid bilayer has been a model system for the biological membrane since the early days of membrane research. Biological membranes show a large compositional heterogeneity and complexity.  They consist of lipid mixtures with varying amounts of unsaturation, different acyl and head groups, and they contain a multitude of imbedded or associated proteins. Lipid bilayers display chain-melting transitions. In the melting regime, the physical state of the membranes is influenced by temperature, pressure, pH, drugs including anesthetics and some neurotransmitters, and the presence of proteins \cite{Heimburg2007a, Heimburg2007c, Seeger2007}.

Electrophysiological membrane models generally assume that the lipid membrane is an insulator and capacitor.  This assumption is important because a significant non-specific ion conductance of the lipid membrane itself would be inconsistent with the specific conductances required by current models of biomembranes. However, membranes are not generally insulators.    Close to their chain melting temperatures, membranes become quite permeable to ions and small molecules \cite{Heimburg2010}.  Temperature changes of only a few degrees have been reported to alter permeation rates for fluorescence markers by several orders of magnitude  \cite{Papahadjopoulos1973, Sabra1996, Blicher2009}. More strikingly, electrophysiological transmembrane current recordings using black lipid membranes (BLMs) or patch pipettes reveal quantized conduction events that resemble those from biological membrane preparations \cite{Yafuso1974, Antonov1980, Antonov1985, Kaufmann1983a, Kaufmann1983b, Gogelein1984, Antonov1985, Antonov2005, Blicher2009, Wodzinska2009, Wunderlich2009} (reviewed in \cite{Heimburg2010}). Antonov and collaborators \cite{Antonov1980, Antonov1985} first described conduction events through membranes made of synthetic lipids eliminating all sources of current fluctuations other than those of the lipid membrane itself.  Other investigations show that recordings of the conductance of synthetic membranes convincingly demonstrate its quantized nature  (eg.,\cite{Blicher2009, Wunderlich2009, Wodzinska2009, Gallaher2010, Heimburg2010}). 

The existence of lipid pores with diameters on the order of 1\,nm has long been discussed in the field of electroporation \cite{Winterhalter1987, Glaser1988, Neumann1999}. Several molecular dynamics studies have demonstrated the generation of such lipid pores by voltage \cite{Tieleman2003, Bockmann2008}. The transient generation of lipid pores by large voltage pulses is used in clinical praxis, e.g. to transport cytostatic drugs into cancer cells or to transfect cells with DNA or RNA segments \cite{Neumann1999, Gehl2003}. It seems likely that the quantized conductance events observed in synthetic membranes are a phenomenon closely related to electroporation. Nevertheless, the quantized nature of the ion currents across lipid membranes is surprising and is not really understood. It suggests that pores or defects in the lipid membrane have a well-defined size. While we believe that the explanation for fixed pore size remains an open issue, the dependence of the pore formation rate on changes in temperature, pressure and other intensive thermodynamic variables is well understood. Its theory is based on the fluctuation-dissipation theorem that treats the couplings of fluctuations in enthalpy, volume, area and other extensive variables with susceptibilities such as heat capacity and compressibility \cite{Heimburg2010}. Thus, the occurrence of the lipid membrane channels responds to changes in temperature, pressure, voltage \cite{Heimburg2010} and (for charged membranes) on pH and Ca$^{2+}$ \cite{Trauble1976}. As a consequence, they are thermosensitive, mechanoreceptive, voltage dependent and pH and Ca$^{2+}$ sensitive. Furthermore, the fluctuation-dissipation theorem implies a connection between the magnitude of the fluctuations and the fluctuation time scale \cite{Grabitz2002, Seeger2007}. Fluctuations in the membrane state are especially large close to melting transitions. Therefore, the mean conductance of the membrane is larger and the channel open lifetimes are longer in the transition regime \cite{Wunderlich2009}. This is particularly important since many biological membranes are in fact found in a state close to a melting transition, e.g., \textit{E.coli} and \textit{bacillus subtilis} membranes, lung surfactant \cite{Ebel2001, Heimburg2005c, Heimburg2007c}, and rat central nerve membranes (unpublished data from 2011 by S. B. Madsen and N. V. Olsen, Copenhagen), suggesting that these phenomena might play a role under physiological conditions.

In this publication we compare the channel events in synthetic lipid membranes with ion channel protein conductances in bio\-membranes. 
The transient receptor potential (TRP) channel family has recently attracted considerable interest due to their involvement in sensing processes. Members of this familiy of ion channels have been reported to respond to environmental stimuli such as temperature, membrane tension, pH, osmolarity, pheromones, and intracellular stimuli such as Ca$^{2+}$ and phosphatidylinositol signal transduction pathways \cite{Wu2010}. They may also be involved in detecting the taste sensations of sweet, sour and umami \cite{Talavera2005, Venkatachalam2007}.

Here, we focus on the activity of three TRP channels, namely TRPA1, TRPM2 and TRPM8. They are considered to form homotetrameric proteins with six transmembrane segments (S1-S6) in each subunit with cytosolic N- and C-termini. TRPA1 is a nociceptive channel and polymodal receptor activated by pungent or irritant chemicals such as AITC (allyl isothiocyanate) from wasabi, acrolein (in smoke), allicin and diallyl disulfide from garlic \cite{Wu2010, DelCamino2010, Doerner2007, Banke2010}. There is evidence for a cold-sensitivity of TRPA1 \cite{Venkatachalam2007, DelCamino2010}. Furthermore, it is sensitive to depolarisation and to Ca$^{2+}$ \cite{Doerner2007}. Pore properties of TRPA1 vary dynamically depending on the presence of Ca$^{2+}$ and agonist stimulation \cite{Banke2010}. The TRPM2 channel is a member of the transient receptor potential melastatin family that was first identified in cancer cells. TRPM2 currents show a linear I-V relationship indicating that this channel is not voltage-dependent. It is activated by intracellular ADP-ribose and Ca$^{2+}$ in a synergistic manner \cite{Csanady2009}, by heat, and by hydrogen peroxide, and this points to a role of the channel in oxidative stress signaling cascades \cite{Bari2009, Wu2010}. A single channel conductance of $\sim$62 pS has been reported for TRPM2 \cite{Wu2010}. The closely related TRPM8 channel is a cold receptor \cite{McKemy2002, Peier2002} and a voltage-gated channel showing strong outward rectification. It is activated by substances that cause the sensation of cold such as menthol or menthol derivatives (e.g. WS-12). The ``super-cooling'' agent icilin is not structurally related to menthol activates TRPM8 in the presence of Ca$^{2+}$  \cite{Chuang2004}. TRPM8 channel activity is modulated by pH \cite{Andersson2004} and the presence of polyunsaturated fatty acids and lysophospholipids \cite{Andersson2007}. A single channel conductance of $\sim$ 81 pS was reported for TRPM8 \cite{Wu2010}.

All three of these channels are selective for cations but show little selectivity for particular cations. Moreover, both TRP channels and lipid membrane channels are influenced by chan\-ges in intensive thermodynamical variables such as temperature and membrane tension (lateral pressure). Therefore, it is tempting to compare the characteristic properties of lipid and protein channel activity. In the present paper we discuss the similarities and differences of TRP channel conductance and conduction events due to synthetic lipid membrane pores. Since menthol is an agonist for human TRPM8 and TRPA1, we also study the influence of menthol on lipid phase behavior. The underlying question is whether membrane pores and protein channels are related or synergetic, and whether it is possible that they are governed by the same physical laws.


\section*{Materials and Methods}
\subsection*{Synthetic membranes}
\textbf{Calorimetry:} Heat capacity profiles were recorded using a VP-DSC (MicroCal, Northhampton/MA, USA) with a scan rate of 5$^\circ$/h.\\
\textbf{Electrophysiological recordings on synthetic lipid membra\-nes:}
We used the droplet method, whereby the planar lipid bilayer membranes is formed on the tip of a patch-clamp glass pipette that has been filled with
electrolyte solution \cite{Hanke1984}. The tip is in contact with the surface of a beaker filled with the same electrolyte solution. Lipids are dissolved in a hexane/ethanol mixture and are then brought into contact with the outer surface of the glass pipette. When the solution flows down the pipette, a membrane is formed spontaneously at the tip of the pipette. The solvent was allowed to evaporate for at least 30 seconds before the pipette was lowered 2--5\,mm below the bath surface. The main advantage of this method is that the resulting membrane is practically solvent free.

Pipettes were pulled from 1.5\,mm / 0.84\,mm (outer diameter / inner diameter) borosilicate glass capillaries (\#1B150F-3, \linebreak World Precision Instruments, USA) with a vertical PC-10 puller from Narishige Group, Japan. A two-step pulling procedure was used, where the first pull was 8mm and the heater was set to 80\% of the instrument's maximum output. For the second pull, the heating coil was lowered 4mm and the heater setting was reduced to 45\%. This produced fairly short and thick pipettes with an hourglass-like taper. For some experiments the pipette was subsequently fire polished using a Narishige MF-900 Microforge. This created pipettes that had a pipette opening in the range from 5--15\,$\mu$m in diameter. As a general rule, larger openings made it more difficult to create a stable membrane, with 20\,$\mu$m being the practical upper limit. In most experiments it was chosen to use the pipette as-is without fire polishing. These pipettes had openings of less than a 1\,$\mu$m.  Unpolished pipettes were use in order to minimize the variation between experiments. The electrodes were made of high-purity, chlorinated silver wires, which were frequently re-chlorinated to avoid baseline drift and additional noise. Pipettes were always freshly prepared immediately before use. The buffers in the pipette and in the medium were identical.

Current recordings were made using an Axopatch 200B patch clamp amplifier (Axon Instruments Inc., Union City/CA, USA). The pipette and electrodes were mounted on a cooled capacitor feedback integrating headstage amplifier (Headstage CV 203BU, Axon Instruments Inc.). The headstage itself was moun\-ted on a micro-manipulator (model SM1, Luigs and Neumann, Germany), allowing for careful and precise control of the pipette position relative to the bath surface. Lastly, the headstage and micro-manipulators were wrapped in a finely meshed metal cloth that acted as a Faraday cage. Data traces were recorded with Clampex 9.2 software (Axon Instruments) using the Whole Cell (headstage gain, $\beta$ = 1), voltage clamp mode. The sampling frequency was either 10kHz or 20kHz, and the signal was filtered by the patch clamp amplifier's analog 4-pole lowpass Bessel filter with the cut-off (-3dB) frequency set to 2kHz.

\subsection*{Cell membranes}
\textbf{Molecular biology and cell culture:}
The cDNAs of human TRPM2, TRPM8 and TRPA1 were subcloned into pIRES-\linebreak hrGFP-2a vectors (Stratagene, USA). Wild-type channels were stably expressed in HEK-293 cells (ATCC,USA) as described previously \cite{Kuhn2010}.\\
\textbf{Solutions:}
Standard bath   solution  (BP1)  contained   140mM  NaCl,  1.2mM  MgCl$_2$ ,  1.2mM CaCl$_2$,  5mM  KCl,  10mM  HE\-PES,  pH  7.4 (NaOH).  Pipette solution contained 145mM cesium glutamate, 8mM NaCl,  2mM MgCl$_2$, 10mM HEPES,  pH 7.2 (CsOH)  and  the Ca$^{2+}$  concentration was adjusted to either $<$10 nM (P10, 10 mM Cs-EGTA),  or to 1 $\mu$M (P10*) as calculated  with  the  MAXC-program  (0.886mM  CaCl$_2$   and 1 mM Cs-EGTA). For  the  stimulation  of TRPM2  currents  in the  inside-out  configuration,  ADPR  (Sigma, 100 mM stock solution  in distilled water)  was added  to  the  intracellular (bath) solution  containing  1 $\mu$M Ca$^{2+}$ , yielding a final concentration of 0.1 mM  ADPR. Alternatively, TRPM2  currents  were evoked  in cell- attached configuration  by application of hydrogen  peroxide (Merck, 30\% stock solution)  to the bath  solution,  and  after  the  appearance  of single  channel  openings, a small membrane patch was excised by lifting up the  pipette  to reach inside-out  configuration.   TRPM8  currents  were induced  with  icilin (Cayman Chemical,  10 mM stock solution  in DMSO) or WS-12 (Tocris, 30 mM stock solution  in DMSO)  by application to the  bath  (final  concentrations as indicated  in the experiments). For stimulation of TRPA1, allylisothiocyanate (Sigma, 30 mM stock solution in DMSO)  was added  directly  to  the  bath  solution. All chemicals  were tested  on  native and vector-transfected HEK-293 cells and did not evoke single channel  currents  in the given range of concentrations without  the presence of a TRP  channel  protein.\\
\begin{figure*}[ht!]
    \begin{center}
	\includegraphics[width=16.5cm]{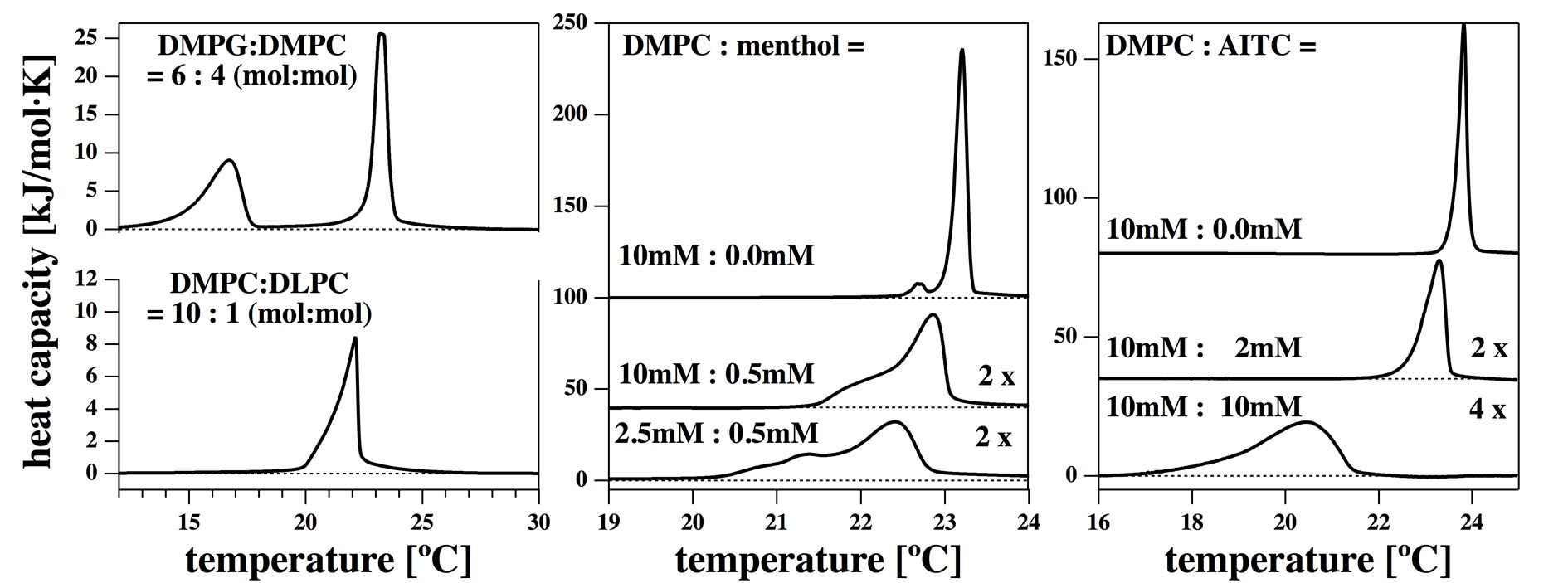}
	\parbox[c]{16cm}{\caption{\small\textit{Left: Heat capacity profiles of multilayered vesicles of a DMPC:DLPC=10:1 mixture in 150mM NaCl (1mM EDTA, 2mM HEPES, pH 7.4) and DMPG:DMPC=6:4 vesicles in 100 mM NaCl (1mM EDTA, 10mM HEPES, pH 7.4). The DMPC:DLPC mixture displays a pronounced maximum at 22.1 $^\circ$C with wings towards higher and lower temperatures. The DMPG:DMPC mixture displays a peak at 23.2 $^\circ$C and a broad pretransition peak at 16.7  $^\circ$C. Center: DMPC membranes in the absence and presence of menthol (150mM KCl, 3mM EDTA, 3mM Hepes, pH 7.4). Different molar ratios are shown: 10mM DMPC in the absence of menthol (top), 10mM DMPC in the presence of 0.5mM menthol and 2.5mM DMPC in the presence of 0.5mM menthol in the buffer (before mixing). One recognizes that the presence of menthol shifts melting profiles towards lower temperatures and broadens the profiles. The bottom two traces have been amplified by a factor of 2 and baselines been shifted for better visibility. Right: 10mM DMPC membranes in the absence (top) and presence of 2mM and 10 mM AITC in the buffer (before mixing). As in the menthol experiments, the presence of AITC broadens and shifts the heat capacity profiles towards lower temperatures. The bottom two traces have been amplified by a factor of 2 and 4, respectively. }
	\label{Figure1}}}
    \end{center}
\end{figure*}
\textbf{Electrophysiology:} 
Single-channel recordings were routinely performed with pipettes made of borosilicate glass (outer diameter 1.8 mm / inner diameter 1.08 mm / length 75 mm, Hilgenberg, Germany). Alternatively, microelectrodes were made from borosilicate glass with slightly different specifications (outer diameter 1.5 mm / inner diameter 0.86 mm / length 100 mm, Harvard Apparatus, USA) to ensure that biophysical properties of ion channels were not dependent on the type of pipette used. Pipettes were fabricated in a two-step procedure and firepolished with a programmable  DMZ-Universal puller (Zeitz-Instrumente GmbH, Germany). Pipettes had a  tip diameter in the range of 1 $\mu$m and resistances between 5 and 7 M$\Omega$. To reduce thermal noise, pipette tips were coated with dental wax (Moyco Industries Inc., USA). Microelectrodes were used on the day of preparation and stored in a container for the prevention of dust deposits. \\
Electrophysiological signals were recorded with an Axopatch 200B amplifier in combination with  Digidata 1440A AD/DA converter controlled by the pClamp10 software suite (Axon CNS, USA). The CV 203 BU headstage (Axon CNS, USA)  was mounted on a micro-manipulator (model SM1, Luigs and Neumann, Germany) inside a Faraday cage. Transfected cells were visually identified with an Axiovert 200 inverted microscope (Carl Zeiss MicroImaging GmbH, Germany) and a blue LED lamp (model CREE XP-E, wavelength $\lambda$ =  460 nm) serving as a light source for excitation of green fluorescence from GFP (green fluorescent protein). The Axopatch 200B amplifier was run in resistive whole-cell voltage-clamp mode ($\beta$=1) to avoid contamination of traces with reset glitches due to capacitor discharge. The output gain ($\alpha$) was set to $\times$100 when recording single-channel currents. A gap-free acquisition mode was used with a sample rate of 10 or 20 kHz and analogous filtering at 2 or 5 kHz performed with a build-in 4-pole Bessel filter (-3 db). All experiments were done at room temperature (21$^{\circ}$C-23$^{\circ}$C). Voltage signals were not corrected for liquid junction potentials.    In order to facilitate comparison of lipid and protein traces, current and voltage signals are depicted as recorded with Clampex software from pClamp10 program suite. It is common in physiology to invert single-channel currents recorded in the cell-attached and inside-out configuration so that the inward movement of ions is represented as downward deflection. The reason for this lies in the convention that the net movement of positive ions in the direction of the outer to the inner membrane is, by definition, an inward current, but as positive ions are leaving the headstage and patch pipette this would be recorded as positive (or upward) current \cite{AxonGuide2008, Wyllie2007}. The patch-clamp command voltage is positive if it increases the potential inside the micropipette. In physiology, it is common usage to report the transmembrane potential (V$_m$), i.e. the potential at the inside of the cell minus the potential at the outside instead of pipette potential. In cell-attached and inside-out configuration where the pipette is connected to the outside of the membrane, a positive command voltage causes the transmembrane potential to become more negative, therefore it is hyperpolarizing \cite{AxonGuide2008}. In the inside-out and cell-attached configuration, the transmembrane potential is inversely proportional to the command potential, in cell-attached configuration V$_m$ is additionally shifted by the resting membrane potential of the cell \cite{AxonGuide2008, Wyllie2007}.

\section*{Results}
The results section contains data regarding the thermodynamic properties, channel-like ion conductance events, and current-voltage relationships of synthetic lipid membranes.  We then compare conductance events from the synthetic membranes with biological membranes containing various transient receptor potential (TRP) ion channels. \\

Fig. \ref{Figure1} shows the heat capacity profiles of two lipid mixtures, DMPC:DLPC=10:1 and DMPG:DMPC = 6:4 in a 150 mM NaCl or KCl buffer. We have used these two mixtures for the conductance recordings that are documented below. Both mixtures display a maximum close to room temperature where our conductance studies were performed. In the transition, the compressibility of the membrane is at maximum \cite{Heimburg1998, Ebel2001}. This effect results in a maximum of the membrane permeability close to the transition temperature \cite{Papahadjopoulos1973, Nagle1978b} and the maximum probability of finding lipid pores or lipid ion channel events \cite{Blicher2009, Wunderlich2009}. This effect is central to our description in the following figures. Menthol is an agonist of TRP channel activity. For this reason we investigated the influence of menthol on lipid membrane phase behavior. The center panel of Fig. \ref{Figure1} shows the melting profile of DMPC membranes with and without menthol. The right hand panel shows the same membrane in the presence of AITC. AITC is an activator of the TRPA1 channel. It can be seen that small amounts of menthol (500\,$\mu$M) and AITC (2\,mM) have a significant effect on the melting profile both with respect to peak position and width. In particular, menthol lowers the melting peak and broadens the profile significantly. The effect of both menthol and AITC is quite similar to that of a general anesthetic molecule on lipid membranes \cite{Heimburg2007c}. The presence of both menthol and AITC should result in measurable effects on lipid membrane permeability for ions if it is measured close to the transition temperature. 
\begin{figure*}[ht!]
    \begin{center}
	\includegraphics[width=16.5cm]{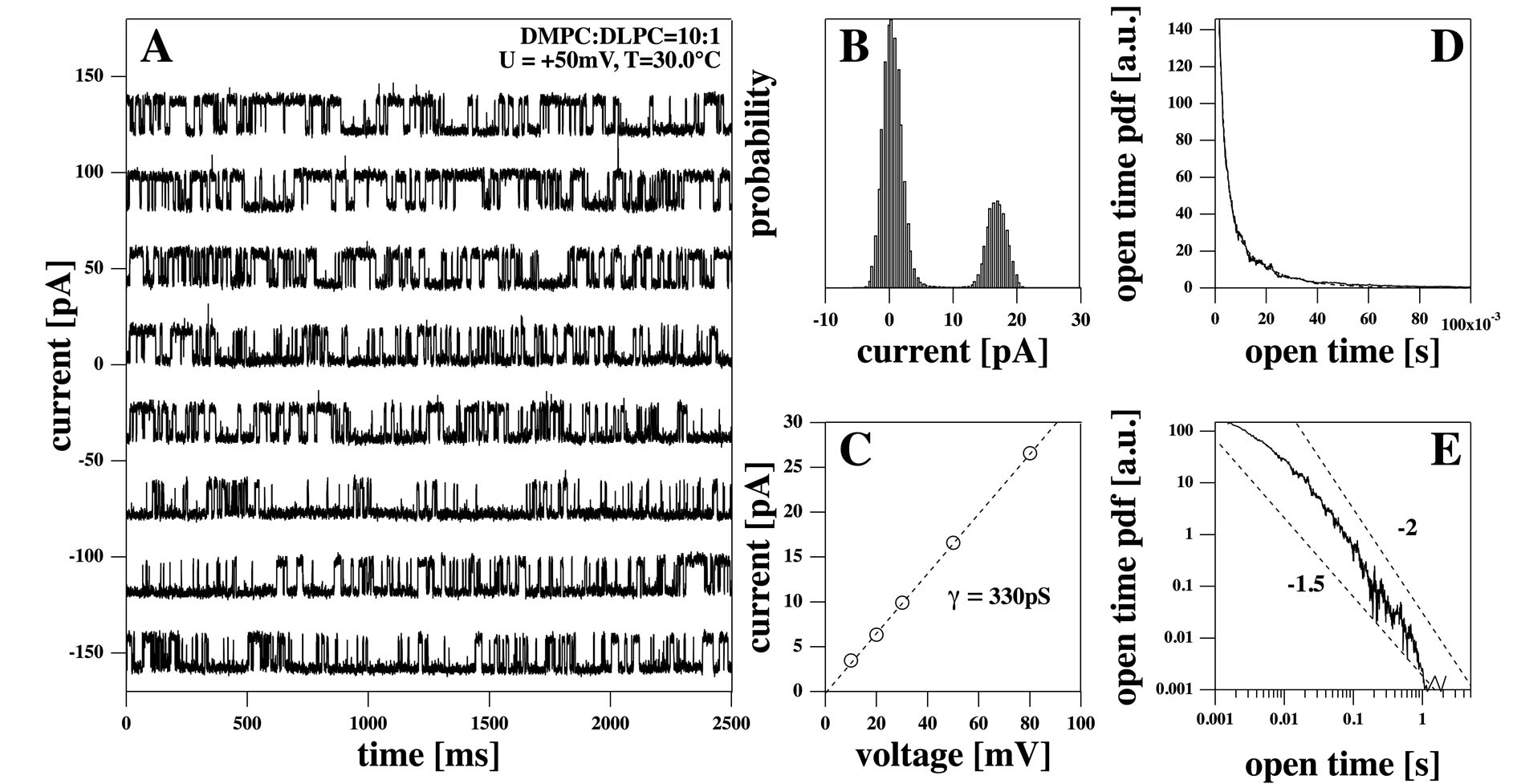}
	\parbox[c]{16cm}{\caption{\small\textit{Patch clamp recording of a synthetic lipid membrane (DMPC:DLPC=10:1 mol:mol,  T=30$^\circ$C, 150mM KCl, 1mM EDTA, 2mM HEPES, pH 7.4) at U=+50\,mV. Panel A:  8 representative 2.5 second segments out of a 30 minute recording. Panel B: Current histogram for the quantized steps of the left hand panel (closed channel state set to zero). Panel C: current-voltage relation for the currents of the membrane shown in the left hand panel (only positive voltages were recorded). The channel conductance is about  330pS. Panel D: Probability distribution function (pdf) for the open times of the channels in panel A. It is well fitted by an biexponential profile in the range up to 100ms open time. Panel E: The double-logarithmic representation of the data in panel D shows that for long open times the pdf is better described by a power law with a critical exponent close to -2.}
	\label{Figure1a}}}
    \end{center}
\end{figure*}

Fig. \ref{Figure1a}A shows a channel recording of a DMPC:DLPC=10:1 mixture in 150\,mM KCl at 50\,mV at  30$^\circ$C. Here and in all following experiments on synthetic membranes, the membrane was deposited on a patch pipette tip using a method described by \cite{Hanke1984}. This trace was stable over the complete recording time (about 30 minutes). The current histogram shows two well-defined peaks for baseline and single channel opening with an amplitude of about 18\,pA at 50\,mV (Fig. \ref{Figure1a}B). Also shown is the linear current-voltage relation of the single channel currents with the baseline subtracted. The  single channel conductance resulted in a  conductance of about 330 pS (Fig. \ref{Figure1a}C).  No single-channel events were detected for negative voltages. Thus, the single-channel current-voltage relationship is only shown for positive voltages.  Panels D and E show the probability distribution function (pdf) for the channel open times in linear and double-logarithmic representation. In the linear representation (Fig. \ref{Figure1a}D), it seems as if the pdf is well fitted by a bi-exponential curve with characteristic open lifetimes of 2.5\,ms and 14.4\,ms (the fit in the range up to 100\,ms is shown as a dashed line that is hardly visible because it overlaps the pdf-profile).  In the double-logarithmic representation, however, it is obvious that the bi-exponential fit underestimates the long open time probabilities, and the long open times are better described by a power law with a fractal exponent between 1.5 and 2  (Fig. \ref{Figure1a}E). Similar power-law behavior was found for the closed time pdf (data not shown). 
\begin{figure}[ht!]
    \begin{center}
	\includegraphics[width=8.5cm]{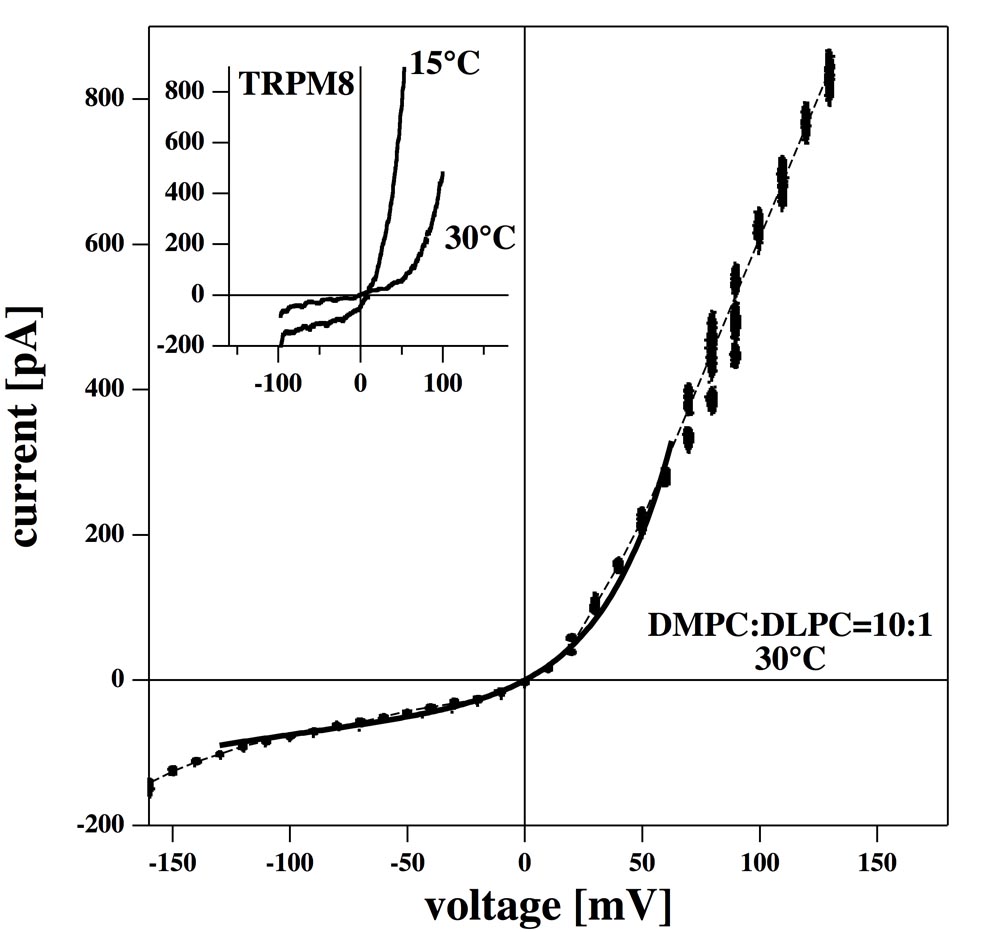}
	\parbox[c]{8cm}{\caption{\small\textit{The current-voltage relationship of the total membrane (DMPC:DLPC=10:1 mol:mol, T=30$^\circ$C, 150mM NaCl, 1mM EDTA, 2mM HEPES, pH 7.4) shows an outward rectification that is probably due to asymmetries of the patch setup (e.g. induced by slight suction or by pipette shape). The dotted line is a guide to the eye. The solid line is a fit for the Eyring transition state model using a barrier position of $\delta=0.86$ and equal ion concentration on both sides of the membrane. The fit was from -130\,mV to 70\,mV. The insert shows the outward rectified profiles of the TRPM8 channel at two temperatures (data adapted from \cite{Voets2004}).
	}
	\label{Figure1c}}}
    \end{center}
\end{figure}

As single channel events we consider stepwise conductance changes on top of a baseline. While the baseline current is usually believed to be related to leaks introduced by bad seals, this is not what is typically found for the synthetic membranes. We have consistently found that the background conductance reflects the thermodynamic properties of the membrane. For instance, the conductance of the overall membrane reflects the heat capacity profile \cite{Wunderlich2009}. The conductance of single channels is strictly linear in the experiments presented here but channel open likelihoods increase with voltage \cite{Wodzinska2009} leading to a non-linear mean conductance. Within a certain voltage range, the current-voltage relation of the baseline is also linear. It is possible that the baseline current consists in part of unresolved conduction events that are of shorter duration than the recording resolution and the time constant of the low pass filter. In Montal-M\"uller type black lipid membranes, we have often found non-linear but symmetric current voltage relationships. This is to be expected for a fully symmetric bilayer. An example is given in \cite{Blicher2009, Wodzinska2009} for a DOPC:DPPC= 2:1 mixture (150 mM KCl, pH = 6.5 at T =19$^\circ$C). 
\begin{figure}[ht!]
    \begin{center}
	\includegraphics[width=9cm]{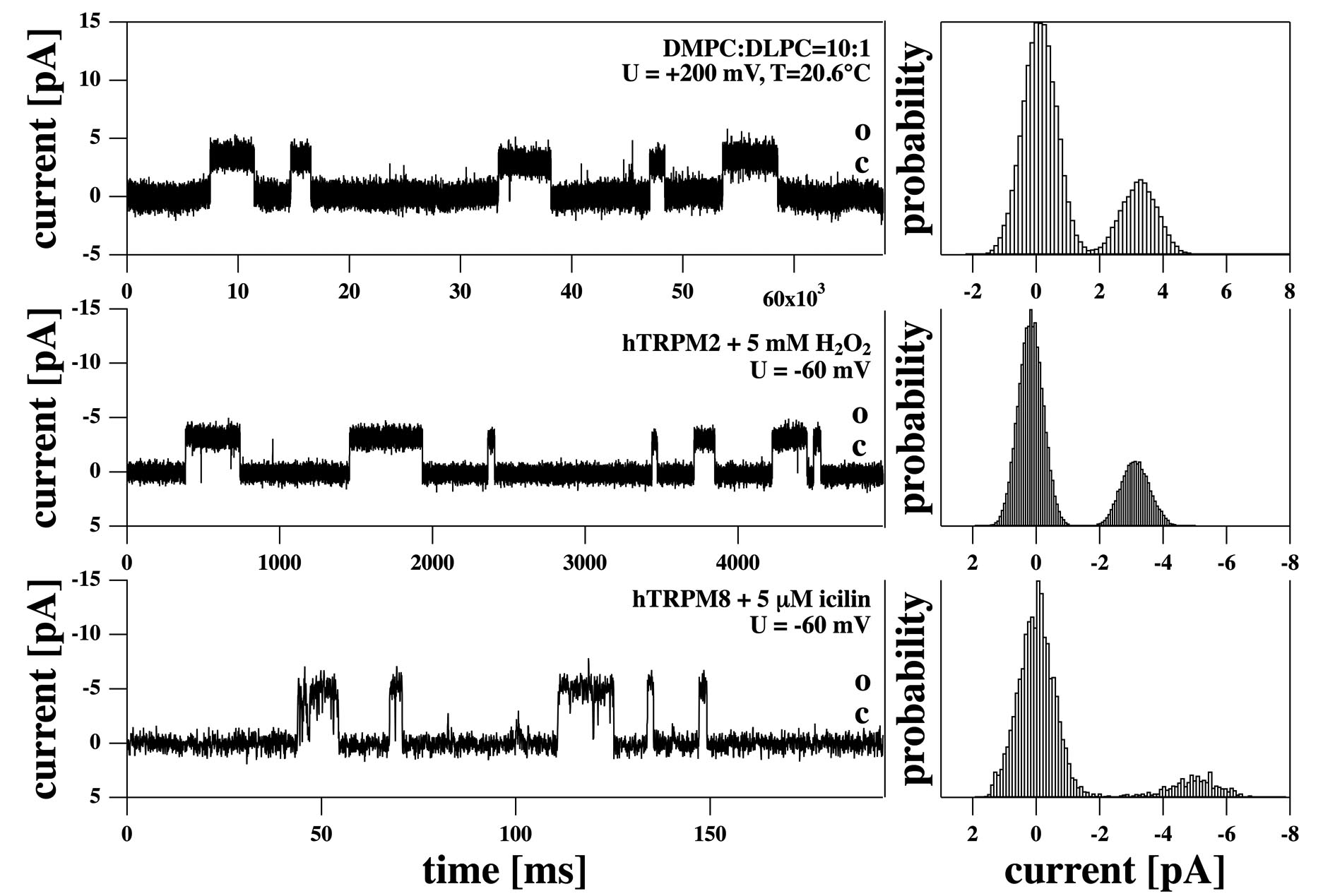}
	\parbox[c]{8cm}{\caption{\small\textit{Single channel recordings from synthetic lipid membranes and from two TRP channels with the corresponding histograms. Top: DMPC:DLPC=10:1, 150mM KCl, T=20.6$^\circ$C and U = +200\,mV with a current of about 3.5pA. Center: hTRPM2 channel activated by H$_2$O$_2$ at U= -60\,mV (pipette voltage, inside-out configuration, pipette solution P10$^\ast$, bath solution BP1 + 5mM H$_2$O$_2$). Bottom: hTRPM8 channel activated by 5 $\mu$M icilin at a pipette voltage of U=-60\,mV (inside-out configuration, pipette solution BP1, bath solution P10$^\ast$ + 5 $\mu$M icilin, low pass filter 5 kHz).
	}
	\label{Figure2a}}}
    \end{center}
\end{figure}

In experiments on patch pipettes we often find non-symme\-tric current-voltage profiles. In Fig. \ref{Figure1c} the average conductance of a membrane with identical composition as in Fig. \ref{Figure1a} (including the baseline conductance) is shown as a function of voltage. We find a nonlinear current-voltage relation with outward rectification shown here over an interval from $-150$\,mV to $+150$\,mV. For comparison, we include an insert showing the current-voltage profiles of the TRPM8 channel at two temperatures which were adapted from \cite{Voets2004}. The I/V relationships from the synthetic membrane and the biological channels show striking similarity regarding their strong outward rectification properties. Since the TRPM8 channel is temperature sensitive, one finds different profiles at different temperatures. Because the mean conductance of lipid membranes also displays temperature dependence \cite{Wunderlich2009}, we expect temperature sensitivity also for the current-voltage relationships.  This issue has not been studied systematically here. It should be added that we  always find outward rectified curves with a significant variation in magnitudes found in individual experiments. We have never seen an inward-rectified profile for any synthetic membrane. This is probably a consequence of the asymmetry of our experimental setup. We cannot exclude that conditions exist for which inward rectification indeed exists.
\begin{figure}[b!]
    \begin{center}
	\includegraphics[width=9cm]{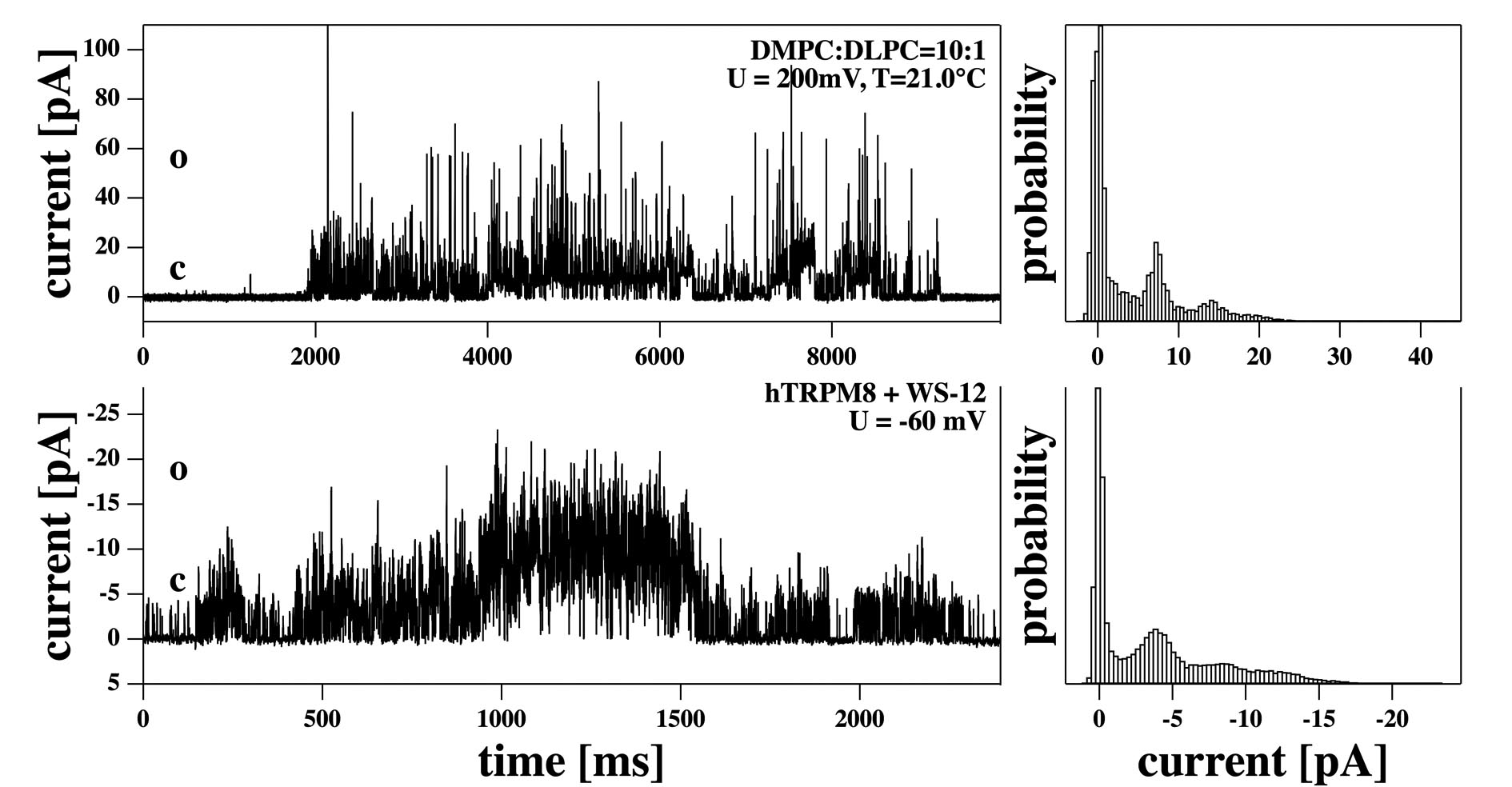}
	\parbox[c]{8cm}{\caption{\small\textit{Conduction bursts in pure synthetic and in a cell membrane with TRP channels, and the corresponding current histograms. Top: DMPC:DLPC=10:1 (150mM KCl, T=20.6$^\circ$C and U = + 200\,mV). Bottom: Burst from hTRPM8 expressed in HEK 293 cells at a pipette potential of U=-60\,mV after addition of the menthol derivative WS-12. Cell-attached configuration, pipette solution P10, bath solution BP1 + 7.5 $\mu$M WS-12.
	}
	\label{Figure2b}}}
    \end{center}
\end{figure}

In the present manuscript we have exclusively used patch pipet\-tes to monitor currents through synthetic membranes. Pi\-pette apertures are much smaller than the holes in the teflon film of BLMs. Fig. \ref{Figure1c} shows a conductivity profile with outward rectification (conductivity increases with increasing voltage) indicating that the membrane is not symmetric. One way  to describe rectified current-voltage relationships is the transition state (Eyring) model \cite{Tsien1969}, which assumes a free energy barrier for the ions, $\Delta G_0$, inside of the membrane \cite{Johnston1995}. The relative position of this barrier within the membrane is given by a parameter $\delta$, with $0\le \delta \le 1$. For a symmetric membrane $\delta$ is equal to 0.5. The electrical potential is assumed to change linearly across the membrane, which means that a uniform dielectric constant within the membrane is assumed. In the presence of an applied voltage, the height of the kinetic barrier has different values, and one finds a current-voltage relationship of the following form:
\begin{eqnarray}
\label{eq:Eyring1}
I&=&z F\beta k_0 \Big[[C]_{in}\exp\left(\frac{\delta F U z}{RT}\right)\nonumber\\
&& - [C]_{out}\exp\left(-\frac{(1-\delta) F U z}{RT}\right)  \Big]
\end{eqnarray}
where $z$ is the charge of the respective ion, $F$ is Faraday's constant, $\beta k_0$ is a rate constant reflecting the height of the barrier and the solubility of ions in the membrane. $[C]_{in}$ and $[C]_{out}$ are the ion concentrations inside and outside of the membrane. For equal ion concentration outside and inside ($[C]_{in}=[C]_{out}$), the current $I$ is zero at $U=0$, as measured in Fig. \ref{Figure1c}.
For a monovalent salt ($z=1$) eq. \ref{eq:Eyring1} simplifies to 
\begin{equation}
\label{eq:Eyring2}
I \propto \left[\exp\left(\frac{\delta F U}{RT}\right) - \exp\left(-\frac{(1-\delta) F U}{RT}\right)  \right]
\end{equation}
This model describes the movement of gating particles in the Hodgkin-Huxley gate model for channel proteins \cite{Johnston1995}. It has also been used to describe the voltage-sensitive conductivity of membrane patches containing TRP channels \cite{Nilius2005}. If the barrier is placed at a position $\delta=0.5$ (the center of the membrane), the current-voltage relationship is symmetric under sign changes of the voltage.  The transition state model is expected to work only if the height of the barrier is greater than the free energy of the ion in the field ($\Delta G_0 \ge |\delta F U z|$ and $\Delta G_0 \ge |(1-\delta) F U z|$, see the derivation of eq. \ref{eq:Eyring1} in \cite{Johnston1995}), indicating that even under idealized conditions the relationship can only be used within a certain voltage regime. In Fig. \ref{Figure1c} we show a fit of the total membrane conductance over the voltage range of $-130$\,mV to $+70$\,mV using a barrier position of $\delta = 0.86$ that describes the conductance of the membrane reasonably well. The position of the barrier at $\delta=0.86$ indicates that the membrane is asymmetric.  Although we cannot offer a conclusive explanation of this result, a number of possibilities present themselves.  One is the application of suction in our patch experiments, which can lead to a membrane curvature that could contribute to the current-voltage asymmetry. The patch pipette itself introduces an asymmetry since it contacts only one side of the membrane.  It is known that the contact of membranes with glass surfaces influences their phase behavior \cite{Yang2000}.  This effect is of little relevance in the whole-cell patch-clamp configuration which was employed to record the current-voltage relationship of TRPM8 depicted in the insert of Fig. \ref{Figure1c}. In whole-cell configuration, the membrane is ruptured under the patch providing access to the intracellular space of the cell. The pipette contacts only a small area of the membrane surface while recording the current flowing over the membrane of the entire cell.\\
\begin{figure}[t!]
    \begin{center}
	\includegraphics[width=9cm]{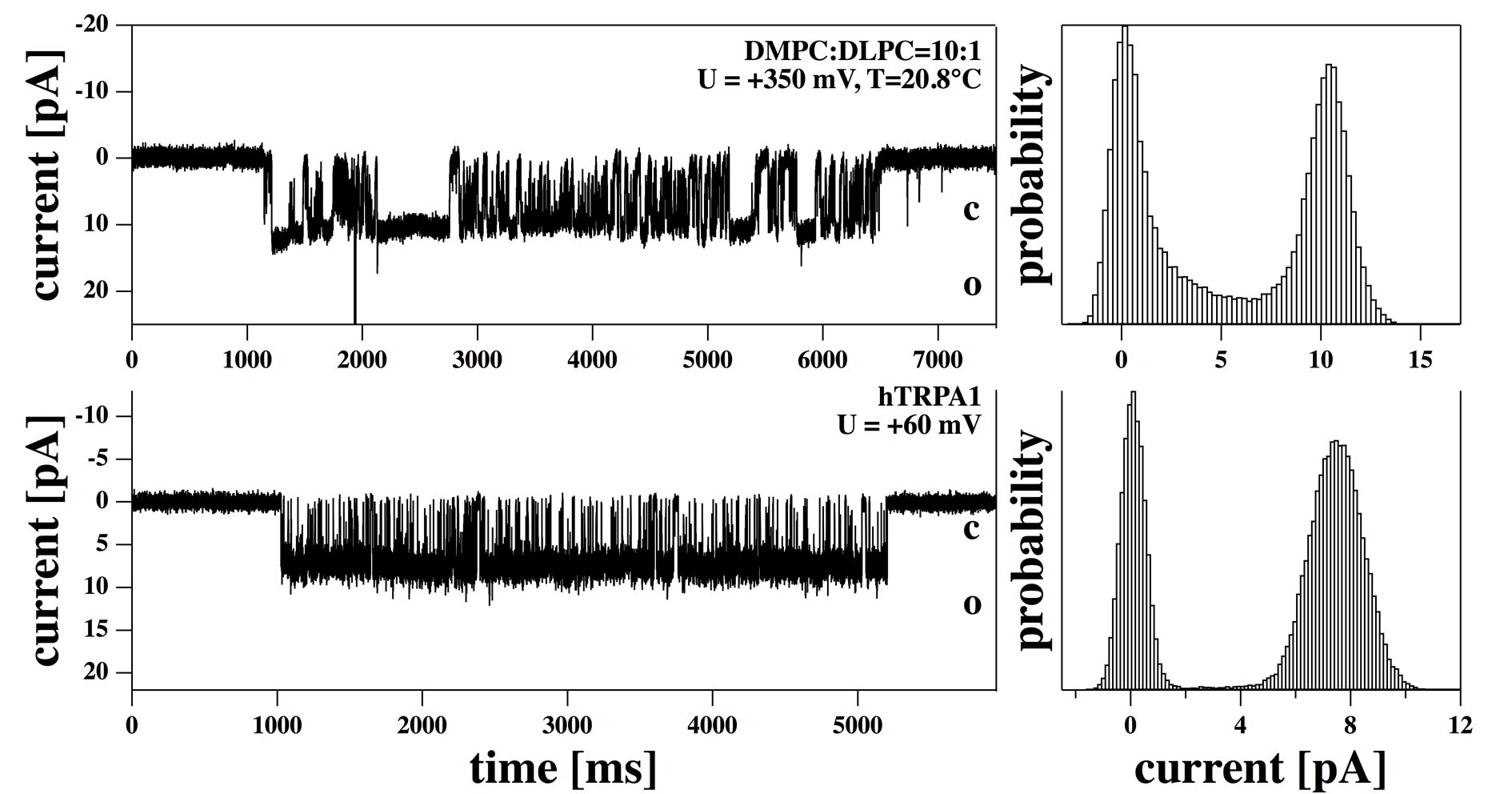}
	\parbox[c]{8cm}{\caption{\small\textit{Flickering of a synthetic and a cell membrane containing TRP channels. Top: DMPC:DLPC=10:1 (150mM KCl, T=20.8$^\circ$C and U = +350\,mV). Bottom: hTRPA1 in HEK293 cell in cell attached configuration at a pipette potential of U= +60\,mV. Pipette solution BP1, bath solution P10.
	}
	\label{Figure2c}}}
    \end{center}
\end{figure}

\noindent\textbf{Comparison of current recordings from synthetic lipid membranes and biological preparations containing TRP channels:}
The current recordings from artificial membranes showed various behaviors including single channel openings, conductance bursts, and flickering.  For some samples and temperatures we found several of these phenomena in the same recordings. The occurrence of these lipid channel events differed somewhat for different pipettes and different conditions.  In contrast, TRP channel activity was independent on the type of pipette used and reproducible under experimental conditions. In the following we compare some typical findings selected from artificial membranes with recordings from HEK293 cells containing TRP channels. All voltages are given as pipette potentials.

Fig. \ref{Figure2a} (top) shows single channel events in a DMPC:\linebreak DLPC=10:1 membrane with single steps of about 3.3\,pA at a voltage of $+200$\,mV (corresponding to $\sim$17\,pS) at a temperature of 20.6$^\circ$ C. This experiment shows that one finds single current events not only at 330\,pS (Fig. \ref{Figure1a} recorded at 30 $^\circ$C) but also much smaller values depending on conditions. We compare this with the current from a HEK293 cell with over-expressed hTRPM2 channels activated by 5\,mM H$_2$O$_2$ at a voltage of $-60$\,mV in inside-out configuration (Fig. \ref{Figure2a}, center). We found single steps of about 3.1\,pA corresponding to a conductance of 52\,pS.   Fig. \ref{Figure2a} (bottom) shows the hTRPM8 channel in a HEK293 cell activated by 5\,$\mu$M icilin at a voltage of $-60$\,mV in inside-out configuration. Single-channel openings have a current amplitude of 5.1\,pA corresponding to a conductance of 85\,pS. Time axis scaling was adapted for the three traces to depict channel openings with a similar appearance. While the openings of TRPM2 are very long compared to TRPM8, both TRP channels show relatively short openings compared to the above trace from the synthetic membrane. 
\begin{figure}[t!]
    \begin{center}
	\includegraphics[width=9cm]{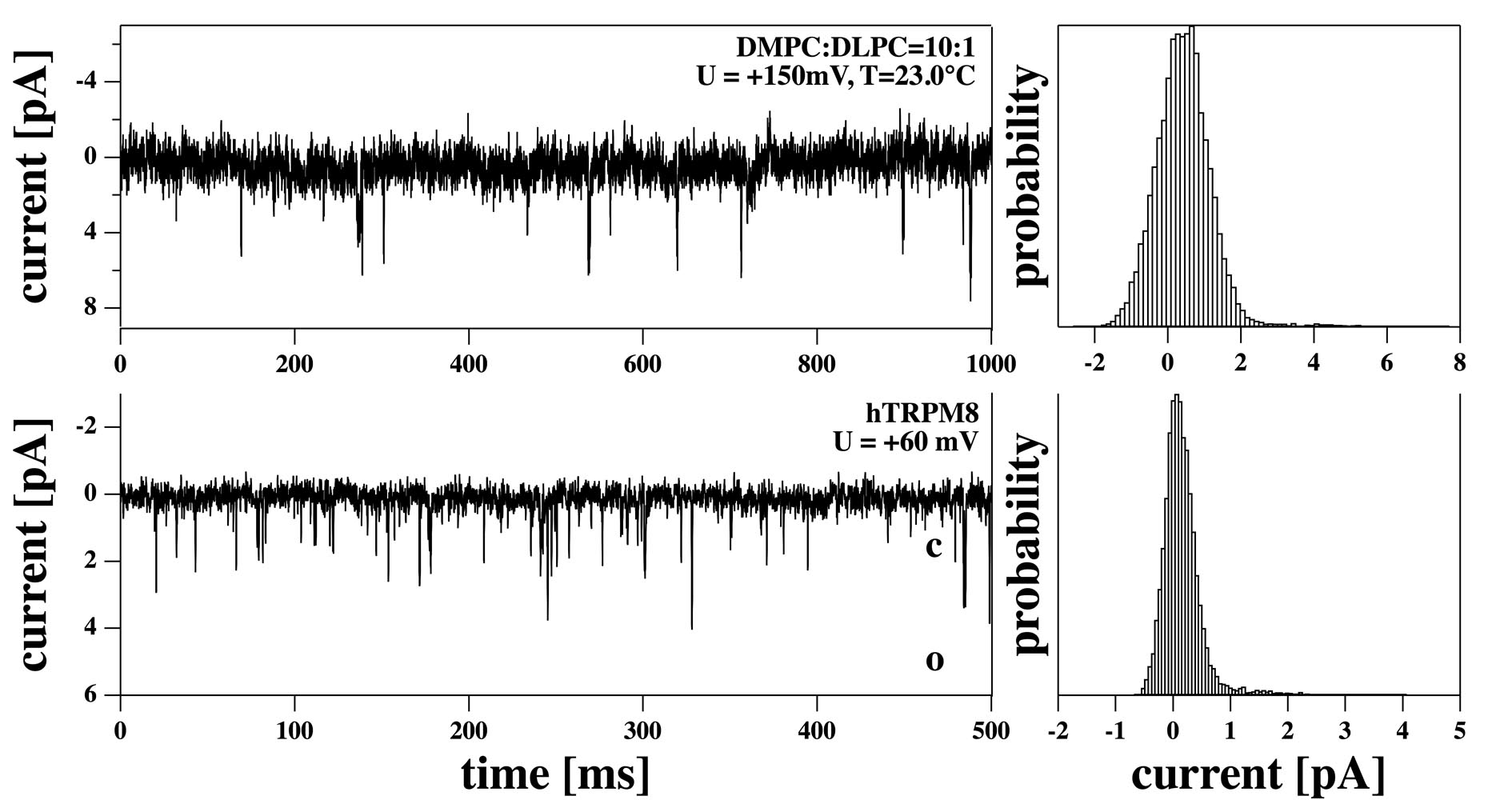}
	\parbox[c]{8cm}{\caption{\small\textit{Spikes from synthetic membranes and from recombinantly expressed TRP channels. Top:  Spikes in a recording from a DMPC:DLPC=10:1 mol:mol membrane (T=23$^\circ$C, 150mM KCl, 1mM EDTA, 2mM HEPES, pH 7.4) at U=+150\,mV. Bottom: Spikes from hTRPM8 in HEK293 cells in cell-attached configuration at a pipette potential of U = +60\,mV. Pipette solution P10, bath solution BP1.
	}
	\label{Figure2d}}}
    \end{center}
\end{figure}
Pe\-riods of repetitive channel activity separated from each other by long closures are called ``bursts''. The conductance events of both synthetic membranes and biomembranes often occur in such bursts. Fig. \ref{Figure2b} (top) shows a burst in a synthetic DMPC:\linebreak DLPC=10:1 membrane recorded at $200$\,mV and T=20.6$^\circ$ C. The histogram indicates approximately evenly spaced conductance levels with an individual step size of 6.8\,pA (34\,pS).  The total burst lasts for about 7 seconds. Fig. \ref{Figure2b} (bottom) shows a burst of an HEK293 cell membrane containing the hTRPM8 channel activated by 7.5\,$\mu$M WS-12 at a voltage of $-60$\,mV in cell attached configuration. The bursting behavior of TRPM8 occured for several minutes after the activator was given, only a short section is shown in Fig. 5. The histogram consists of approximately equally spaced steps of -3.5\,pA (58\,pS). Thus, the bursts of the synthetic membrane and the cell membrane display very similar characteristics both in the conductance of individual steps and in the appearance of the current histogram. Again, the traces by themselves appear nearly indistinguishable in both preparations. 
\begin{figure}[ht!]
    \begin{center}
	\includegraphics[width=9cm]{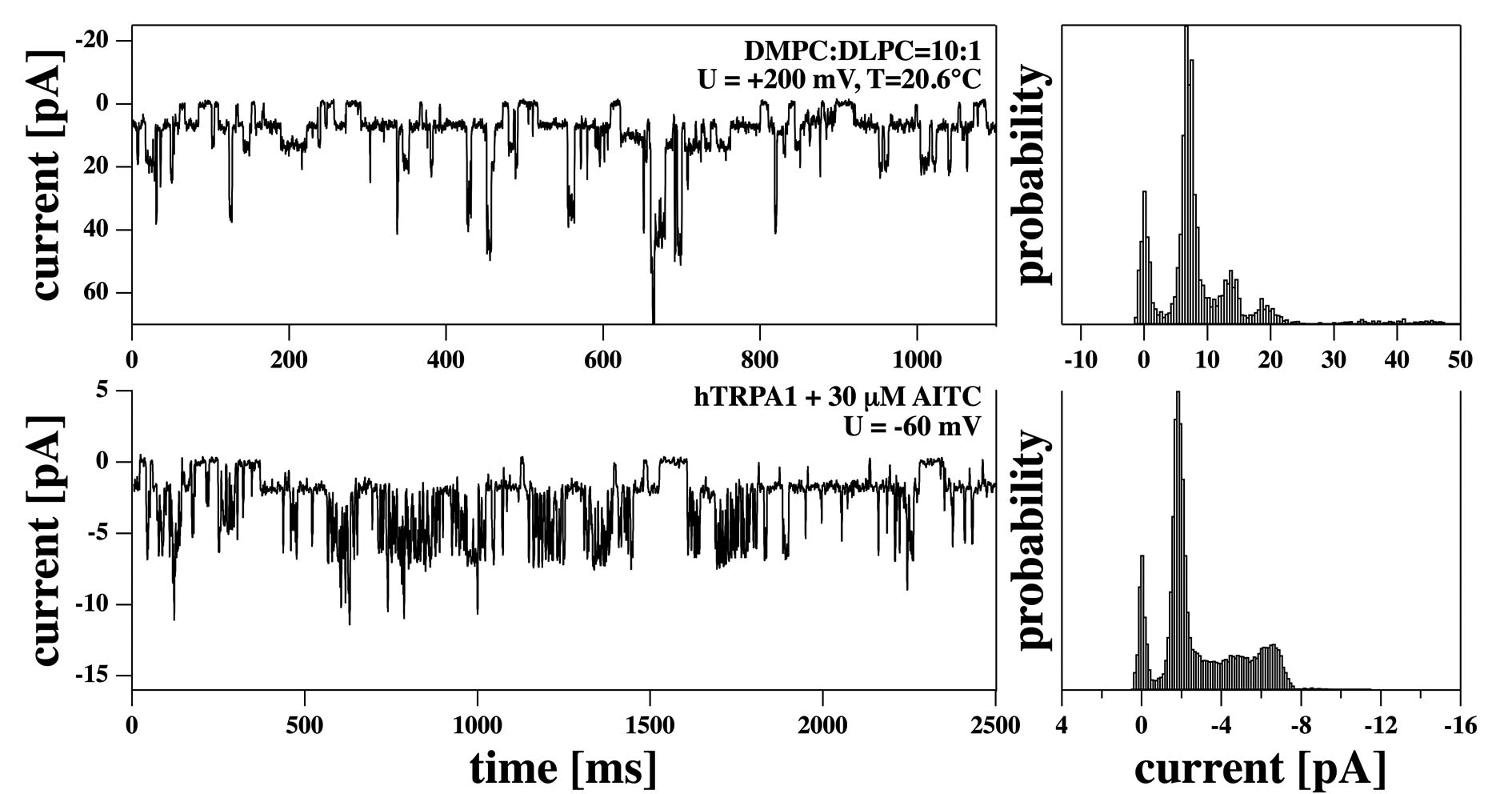}
	\parbox[c]{8cm}{\caption{\small\textit{Multistep conductance in synthetic membranes and in a cell transfected with TRP channels. Top: DMPC:DLPC =10:1 mol:mol membrane (T=20.6$^\circ$C, 150mM KCl, 1mM EDTA, 2mM HEPES, pH 7.4) at U=+200\,mV. Bottom: hTRPA1 in HEK293 cells after addition of AITC in inside-out configuration at U= -60\,mV (pipette voltage). Bath solution P10*, pipette solution BP1, post-recording low pass filtering with 500 Hz. In both traces the lowest conductance level has been set to zero.
	}
	\label{Figure2e}}}
    \end{center}
\end{figure}

Flickering resembles a burst with a single conductance step. Examples of this phenomenon are depicted in (Fig. \ref{Figure2c}). The top trace (Fig. 6, top) shows a flickering event from a synthetic DMPC:DLPC=10:1 membrane, the lower trace (\ref{Figure2c}, bottom) shows the hTRPA1 channel in a HEK293 cell membrane recorded in cell-attached configuration. The flickering event in the artificial membrane occured at 350 mV and 20.8$^{\circ}$C and lasts for about 6 seconds. 
The step size is 10.2\,pA corresponding to a single channel conductance of 23\,pS. 
The current trace of a cell preparation expressing TRPA1 was recorded at 60\,mV and displays a unitary current of 7.4\,pA corresponding to a single-channel conductance of 123\,pS. 
Except for small differences in detail, the overall appearance of the traces does not allow us to distinguish easily between the artificial membrane events from the cell preparation on the basis of the traces alone.

In Fig. \ref{Figure2d} we show channel openings which are so brief that they are mostly not fully resolved, leading to a pattern of ÒspikeÓ-like appearance with variable single-channel amplitudes truncated by low-pass filtering. The top trace is for a DMPC:DLPC=10:1 membrane recorded at 150\,mV and 23$^\circ$\,C and shows spikes with an amplitude of 4--5\,pA corresponding to about 30\,pS. The bottom trace is for a cell-attached recording of an HEK293 cell containing the hTRPM8 channel at U=60\,mV.  Here, the spike has an amplitude of about 2\,pA corresponding to about 33\,pS.  The limited time resolution of the recording system may have lead to an underestimation of the mean current related to the spikes.

Fig.~\ref{Figure2e} shows a comparison of multi-step conductances in synthetic membranes and in a biological preparation containing the hTRPA1 channel activated by 30\,$\mu$M allyl isothiocyanate (AITC). The synthetic membrane display at least four clearly distinguishable and equally spaced conduction levels with a conductance of 34\,pS. 
The histogram of the biological preparation shows four visible steps. Since the two small peaks of the lower histogram are diffuse, two interpretations are feasible: In accordance with the upper histogram, four identical steps may be identified with a mean conductance per step of approximately 28\,pS.
\begin{figure}[ht!]
    \begin{center}
	\includegraphics[width=9cm]{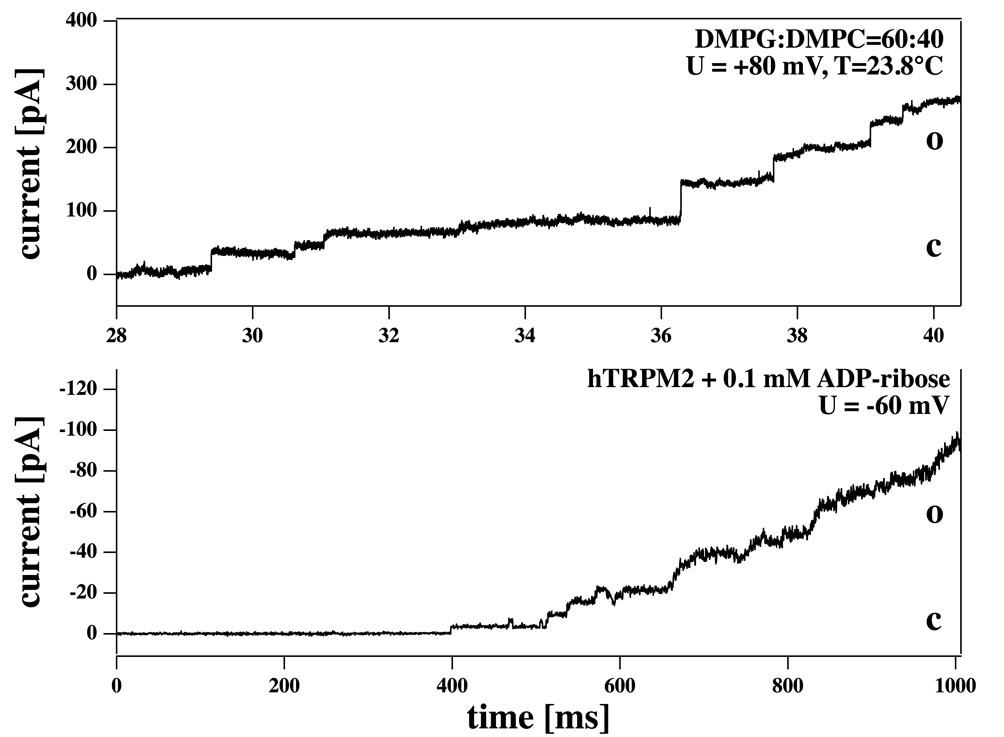}
	\parbox[c]{8cm}{\caption{\small\textit{``Stairs'' (subsequent opening of several channels) in synthetic membranes and in a biological membrane with TRP channels. Top: DMPG:DMPC=60:40 mol:mol membrane (T=23.8$^\circ$C, 50mM KCl, 1mM EDTA, 10mM HEPES, pH 7.4) at U=+80\,mV. Bottom: hTRPM2 channel in  HEK293 cells activated with ADP-ribose at a pipette potential of U= -60\,mV in inside-out configuration. Bath solution P10$^\ast$ + 0.1 mM ADP-ribose, pipette solution BP1.
	}
	\label{Figure2f}}}
    \end{center}
\end{figure}
Alternatively, this pattern of channel activity may be interpreted as long-lived openings of endogenous channels from HEK293 cells \cite{Doerner2007} or a subconductance state of TRPA1 with a small amplitude (2 pA) and overlying larger amplitudes of approximately 5 pA stemming from the main conductance state of TRPA1 with flickery openings \cite{Nagata2005}. The overall appearance of the traces, the life times, and the individual conductances are once again very similar in both systems. While the appearance of the traces is different from the conduction bursts in Fig.~\ref{Figure2b}, the conduction histograms display some similarities both for synthetic and cell membranes.

In practically all synthetic lipid membrane preparations we find channel-like events under appropriate conditions. In particular, when voltage is increased the membranes eventually break at a threshold voltage, probably due to the growth of a very large pore.  Below this threshold one typically finds quantized channel events. The threshold voltage is different in different preparations and depends on temperature, pipette suction, and probably on the properties of the pipette itself.  Increasing the voltage to threshold values leads to rupture visible by a stepwise increase in membrane conductivity as can be seen in Fig. \ref{Figure2f} (top) for a synthetic DMPG:DMPC=6:4 mixture. This onset of the rupture process can be reversed if the voltage is lowered. 
Biological preparations can also display such staircase-like behavior as shown in Fig.~\ref{Figure2f} (bottom) for the TRPM2 channel activated by 0.1mM ADP-ribose in the presence of 1 $\mu$M Ca$^{2+}$. In contrast to membrane rupture, which is mostly an irreversible process, activation of TRPM2 by ADPR is reversible as is obvious from the current going downstairs in a stepwise fashion until finally the baseline is reached (not shown).


\section*{Discussion}
In this paper we have compared conductance traces from biological (HEK293) cells over-expressing TRP channels with recordings from synthetic lipid bilayers. We have demonstrated that suitable adjustments of temperature and voltage (or membrane tension) can always lead to conductance traces in synthetic membranes that are indistinguishable from recordings of biological preparations containing particular proteins. The experimental conditions are not necessarily the same but are comparable to those used in biomembrane experiments. This similarity is sufficiently pronounced that an inspection of short traces may fail  to identify whether recordings are from synthetic  membranes  or from cells.

Protein ion channel activity in patch clamp experiments was characterized by typical and repetitive current events which we\-re visually distinctive among different members of the TRP channel family. Those fingerprint openings from TRP channels with typical conductance and lifetime were absent under control conditions in patches of native HEK293 cells or of vector-transfec\-ted cells. 
In many experiments on pure lipid bilayers, strikingly similar traces were observed, although synthetic membranes did not exhibit the fingerprint-like behavior of channels with their highly reproducible and predictable reactions. Lipid membranes displayed various responses and the conditions for any specific response could not be controlled exactly.  We selected several types of events found in sections of current recordings from artificial membranes for comparison with representative traces of TRP channels pointing phenomenologically at their striking similarities.  We found similar single-channel events, multistep conductance, conduction bursts, flickering, conduction spikes and staircase behavior in synthetic membranes and in HEK293 cells containing three different TRP channels. Conductances and current histograms as well as lifetime distributions are all similar. Further, we found asymmetric non-linear current-voltage relationships in synthetic membranes that could be described by an Eyring-type kinetic barrier model that has also been used to describe the current-voltage relationship of TRP channels \cite{Nilius2005}.  The question arises whether, in spite of their similarity, channel events in the lipid membrane and protein channels can be considered as completely independent. Some findings in the literature suggest that these similarities are not merely accidental and point at a common origin of the conduction events in the two systems. Seeger et al. \cite{Seeger2010} found that both the mean conductance and the open lifetimes of the KcsA potassium channel exactly follow the heat capacity profile of the membrane into which the protein is reconstituted. The protein shows precisely the behavior expected for its host lipid membrane, indicating that lipid membrane properties dominate the experimental characteristics of the system. It was suggested that physical properties of the lipid bilayer influence ion channel activity via a fine-tuning of protein conformation, but it was also pointed out that the lipid membrane itself displays a similar behavior \cite{Seeger2010}.
It was also found that the mean conductance is proportional to the protein concentration. Further, conduction events could still be blocked by tetraethyl ammonium (TEA), which is a KcsA blocker.  Thus, the conduction events are clearly correlated with the presence of the proteins. A similar correlation of channel activity with lipid membrane phase behavior was found for the sarcoplasmic reticulum calcium channel \cite{Cannon2003}, where the highest conduction activity and the longest open times were found close to the phase boundaries of the host lipid matrix. These findings suggest a strong correlation and possibly mutual modulation of the lipid properties with protein channel activity. 

While the importance of protein channels and receptors is wide\-ly accepted, the finding of quantized conductance events in synthetic membranes is striking but little known (e.g., \cite{Yafuso1974, Antonov1980, Antonov1985, Antonov2005, Blicher2009, Wunderlich2009, Wodzinska2009}). Such events occur more frequently close to the melting transition of the lipid membranes. Mean conductances and lifetimes are in agreement with the fluctuation-dissipation theorem that defines the coupling between the amplitude of area fluctuations and the lifetime of the fluctuations \cite{Heimburg2010}. The mean open and closed lifetimes of a lipid membrane channel is closely related to the fluctuation lifetime \cite{Wunderlich2009}. The consequences for lipid membranes can be summarized in the following manner: In the vicinity of the lipid melting transition, area fluctuations and the likelihood of finding pores (or channels) are maximal.  Simultaneously, the fluctuation relaxation time is at maximum as is the mean lifetime of open channels. Different phenomenological behavior as shown in Figs.~\ref{Figure2a}-\ref{Figure2f} probably reflects the properties of different points in the phase diagram of the lipid mixtures. 

Pores in lipid membranes have been discussed for more than 25 years \cite{Papahadjopoulos1973, Nagle1978b, Glaser1988, Tieleman2003, Bockmann2008}. Elastic theory suggests the existence of stable pores of a diameter of approximately 1\,nm. This is close to the suggested size of aqueous pores in ion channel proteins. The experimentally measured pores, i.e., channel events in synthetic membranes, have a diameter similar to that suggested by theory. In Blicher et al. \cite{Blicher2009} aqueous pores of 0.7\,nm diameter were described, close to the theoretical value and close to the aqueous pore sizes suggested for protein channels. 

In our synthetic membrane experiments we typically observe coherent trends. For instance, it is possible to induce channel events in lipid membranes if one increases the voltage above some threshold value. This threshold probably varies with temperature. The same is true for increasing pipette suction.  Further, channel events are more likely in the melting regime.  However, the details of these phenomena varied from experiment to experiment. We believe that this is partially due to strong influences of the patch pipette on the phase behavior of lipid membranes. It is known that the contact of membranes with glass or mica can shift the lipid melting transition by as much as to 5 degrees towards higher temperatures \cite{Yang2000}. For this reason, the melting behavior of our synthetic membrane patches may be influenced by the patch pipette and vary as a function of the exact surface features of the pipette perimeter.  Further, typical domain sizes of lipid mixtures in the transition regime can be several micrometers large \cite{Korlach1999}. With pipette diameters on the order of 1$\mu$m (smaller than some domains) one may therefore observe considerable variation in the relative amounts of fluid and gel domains in different experiments. We have found more reproducible patterns in the larger membrane patches in BLM experiments or in macroscopic permeation events that do not involve interfaces \cite{Blicher2009, Wodzinska2009}. One of the most obvious differences of lipid membrane channels and biological protein preparations is that in the latter case the conductances seem to be more reproducible and stable.  Thus, cells transfected with particular channel proteins display characteristic electrical properties whereas lipid membranes do not exhibit such fingerprint-like features.

In Fig. \ref{Figure1a}E we show that the open time pdf displays power law behavior. The same was found to be true for the closed time pdf (data not shown). Recently we have demonstrated that this applies for other synthetic membranes as well \cite{Gallaher2010}. Power law behavior implies that very long closed times occur with a much higher frequency than in stochastic Markovian kinetics. It leads to the possibility of long silent periods between burst of activity. For protein channels this fractal kinetics has been long discussed \cite{Liebovitch1987a, Liebovitch1987b, Liebovitch1989, Liebovitch2001} as an alternative to multi-state Markovian kinetics \cite{Millhauser1988}.  In this context it is interesting to note that power law behavior is also a natural consequence of critical behavior of membranes close to transitions \cite{Nielsen2000a}. Fig.~\ref{Figure2b} shows that the bursts from TRPM8 in HEK293 cells display very similar phenomenological behavior compared to synthetic membranes close to phase transitions both in respect to channel amplitudes, lifetimes and conduction histograms. This is also true for flickering activity as shown in Fig.~\ref{Figure2c}. Obviously, the temporal patterns of both classes of events obey similar kinetics.

Channels in synthetic lipid membranes can be influenced by drugs if these drugs affect the melting behavior of the membrane. For instance, anesthetics lower and broaden the melting transition of membranes \cite{Heimburg2007c} and thereby have been shown to be able to ``block'' lipid membrane channel events without binding to macromolecular receptors \cite{Blicher2009}. Many other drugs affect the melting behavior and permeability of lipid membranes, e.g.,  the pesticide lindane \cite{Sabra1996} or the neurotransmitter serotonine \cite{Seeger2007}. We have also shown that the anesthetic octanol and serotonine can influence the fluctuation lifetimes of lipid membranes and thereby probably the lipid membrane channel lifetimes. This implies that the response of channels to drugs is not necessarily the consequence of specific binding of the drugs to a receptor. This result could rather be a more general consequence of thermodynamic changes in the surrounding lipid matrix.  Here we have shown that menthol, an agonist of human TRPM8 and TRPA1 channels, and AITC, an agonist of the TRPA1 channel are also able to influence the state of the lipid membrane (Fig.~\ref{Figure1}). In particular, both menthol and AITC lower and broaden the cooperative melting events of membranes in a manner very similar to general anesthetics. This means that menthol and AITC partition in the lipid membrane and are much more soluble in the fluid phase than in the gel phase. In fact, menthol has been compared to general anesthetics (e.g., propofol) \cite{Watt2008}. Interestingly, no specific binding site of menthol is known for TRP channels, in contrast to AITC in the case of TRPA1. For human TRPM8, the amino acids implied in menthol sensitivity are conserved in menthol-insensitive \linebreak TRPM2 \cite{Bandell2006, Voets2007}. This would be in agreement with the idea that the effects of menthol on TRP channels are related to its influence on the host lipid matrix. Above the melting transition, menthol should act as an antagonist of lipid channel formation (because the transition is shifted away from experimental temperature), and below the transition as an agonist (because the transition is shifted towards experimental temperature). Precisely this effect has been observed for general anesthetics such as octanol \cite{Blicher2009}. Interestingly, menthol was reported to have a bimodal action on mouse TRPA1 \cite{Karashima2007}. At high concentrations of menthol an inhibitory effect was found while low concentrations lead to channel activation. In contrast, human TRPA1 is exclusively activated by menthol \cite{Xiao2008}. The authors of this study note that there is no direct evidence that menthol specifically binds to TRPA1, instead the possibility of indirect effects of menthol was considered, e.g. involving modifications of the lipid bilayer. Since effects of menthol on the physical properties of model membranes are better characterized than its binding to TRP channels, it may be speculated that these effects are governed by interaction at the membrane interface.    

The same may apply for the sensitivity to temperature, a hallmark of this family of thermosensitive channels.  We have shown that many biological membranes display transitions \linebreak slightly below body temperature, including E. coli membranes, bacillus subtilis membranes \cite{Heimburg2005c, Heimburg2007c}, bovine lung surfactant \cite{Ebel2001, Heimburg2005c} and the membranes from the central nerve of rat brains (S. B. Madsen and N. V. Olsen, recent unpublished results). Further, we have shown that channel conductance in synthetic lipid membranes displays a strong temperature dependence close to the melting transition of the membranes \cite{Blicher2009, Wunderlich2009}. Transition in biomembranes are found about 10 degrees below physiological temperature \cite{Heimburg2005c} and display a half width of order 10 degrees, corresponding to a van't Hoff enthalpy $\Delta H$ of about 300 kJ/mol and an entropy $\Delta S$ of about 1000 J/mol K. Similar values have been found for the temperature activation of TRP channels. Talavera et al. reported activation enthalpies on the order of 200\,kJ/mol for TRPM4, TRPM5, TRPM8 and TRPV1 channels \cite{Talavera2005}. Further, they report $\Delta S\approx 500$\,J/mol K for several TRP channels. While the origin of the very large $\Delta H$ and $\Delta S$ for the TRP channels remains mysterious, the order of magnitude is just in the range of the melting transition of the biological membrane. For this reason, it has been speculated in the past if the temperature dependence of TRP channels may originate from transitions in the surrounding membrane \cite{Voets2005}.

\section*{Conclusion}
In the present study, we have explored the possibility that lipid bilayers, in addition to their role as electrical  insulators and solvents for membrane  proteins, provide relevant ion permeabilities as well as changes of permeabilities, by themselves and without  the presence of channel-forming proteins.  Indeed, lipid membranes  exhibited  non-linear  and asymmetric current-voltage profiles in pipette  experiments. We  document  a wide spectrum  of electrical  phenomena  in synthetic  membranes such as bursts, spikes, flickering and  multi-step openings that are normally considered typical for the  activity  of protein  ion channels.  Furthermore, we show the  similarity  of current events  from  lipid  bilayers  with  single-channel  recordings  of TRP   channels. Thus,  electrical properties of lipid bilayers may contribute to membrane  excitability generated  by protein  ion channels. On  the  other  hand, electrical  properties of pure  lipid  bilayers  also  display  clear differences as compared  to those  of cell membranes containing  channel proteins, such as the different regulation by specific ligands. Importantly, the  observed  electrical  phenomena  in synthetic  membranes  lack stability  that  would  warrant reproducible responses  in the  presence  of slightly  changing  conditions. This is illustrated  by the problems  to  obtain  strictly  reproducible  electrical  events  with  tip-dipping using glass pipettes on pure lipid bilayers, in contrast to the single-channel  recordings with similar pipettes on cell membranes. \\
The response of lipid membranes to electrical stimuli demonstrated in this study may broaden our view how ion channels in biological membranes are regulated and how channel proteins and their lipid matrix may cooperate in signal transduction in excitable cells.  \\


\noindent\textbf{Acknowledgments:} We thank Prof. Andrew D. Jackson for a careful reading of the manuscript and valuable comments.

\footnotesize{

}

\end{document}